\documentclass[preprint, amsmath, amssymb, showpacs]{revtex4}
\usepackage{bm}
\usepackage[dvipdf]{graphicx}
\DeclareGraphicsExtensions{.jpg,.pdf,.mps,.png,.eps,.ps,.EPS}

\begin{document}
\newcommand{\avg}[1]{\langle{#1}\rangle}
\newcommand{\Avg}[1]{\left\langle{#1}\right\rangle}
\def\be{\begin{equation}}
\def\ee{\end{equation}}
\def\bc{\begin{center}} 
\def\ec{\end{center}}
\def\bea{\begin{eqnarray}}
\def\eea{\end{eqnarray}}
\def\bwt{\begin{widetext}}
\def\ewt{\end{widetext}}
\def\ra{\rightarrow}
\def\ba{\backslash}
\title{Spectral properties  of complex networks}
\author{Ginestra Bianconi}
\affiliation{The Abdus Salam ICTP, Strada Costiera 11, 34014 Trieste, Italy}
%
%

\begin{abstract}
We derive the spectral properties of adjacency matrix of complex
networks and of their Laplacian  by the replica
method combined with a dynamical population algorithm. 
By assuming the order parameter to be a product of Gaussian distributions, 
the present theory provides a solution for the non linear integral equations 
for the spectra density in random matrix theory of the spectra of sparse random matrices making a step forward with respect to the effective medium approximation (EMA) and the single defect approximation (SDA). 
We extend these results also to weighted networks with weight-degree correlations
\end{abstract}

\maketitle
The interest on the spectral properties of complex networks is growing for the 
study of their dynamics
\cite{Doro_critical,Barabasi}
being relevant for example to understand the stability of ecological
networks or  the synchronization stability conditions. \cite{May,Syncrh}
This problem is related with the investigation of spectral properties of random matrices in random matrices ensembles in many fields of theoretical physics. 
\cite{Guh,Bro}
Starting from the discovery of the Wigner semicircle law in 
nuclear spectra \cite{Wig}
the Random Matrix Theory has have a wide range of 
applications  from quantum chaos to irreversible classical 
dynamics and low density liquids \cite{Guh,And,Kra,Cava}

The research on spectral properties of sparse random matrices has started as early 
as in the 1988
\cite{Bray}
but only recently their relevance has been fully 
acknowledged for the study of the properties of a number of dynamical models
defined on the network, like in the models of synchronization.
A number of methods for determining the
density of states of random matrices have been proposed which range from the 
classical results of \cite{Metha} to the replica method formulation and
the supersymmetric formulation.
The works of Monasson and Biroli
\cite{Monasson,Monasson_sw,Monasson_order_pa}
deals with the spectra of Laplacian  matrices of random 
Poissonian networks and small world networks. In \cite{Monasson} a new 
approximation, the so called single defect approximation (SDA) for the study of 
the random spectra has been introduced. The approximation has been further improved 
by the work of Semerjian and Cugliandolo 
 \cite{Semerjian} 
for random Poissonian 
matrices.

Dorogovtsev et al. \cite{Doro_spectra,Doro_critical,Doro_lap}, have developed random 
walk based methods for the evaluation of the spectra of the adjacency matrix and 
the spectra of the Laplacian  of complex networks.

Using the replica method  as in \cite{Monasson,Monasson_sw,Semerjian} the problem reduces to making a good replica
symmetric ansatz for the functional order parameter defined on a
vector of continuous variables defined in the real axis.
In the contest of a statistical mechanics model for studying the
fluxes in the metabolic network \cite{Metabolico,Proc} a similar technical problem was solved
assuming that the functional order parameter can be written as a
weighted sum of Gaussians. The problem was then solved by proposing a population
dynamics to find the statistical weights  corresponding to each
Gaussian in the sum \cite{Metabolico}. 
In this work, following reference  \cite{Metabolico,Proc} 
we will derive the spectra of random matrices using the replica method
and the development of the order parameter in term of weighted Gaussians, a 
technique that  allows also for the extension to weighted matrices of random entries.
This method can be applied both to adjacency matrices and to
Laplacian  matrices providing the tools for the calculation of
different properties of the graph.
\section{Spectra of a matrix}
Given a random matrix of  eigenvalues $\lambda_n$ with $n=1\dots N$,
the spectral density $\rho(\lambda)$ is defined as
\be
\rho(\lambda)=\frac{1}{N}\sum_n \delta(\lambda-\lambda_n)
\ee
and can also be expressed as 
\be
\rho(\lambda)=-\frac{1}{\pi N}\mbox{Im}
\mbox{Tr}\frac{1}{\lambda+i\epsilon-A}
\ee
We suppose that  in the thermodynamics limit the spectral density
is self-averaging, i.e. 
\be
\rho(\lambda)\rightarrow \Avg{\rho(\lambda)}
\ee
where the average is performed over all matrices in a given ensemble.
To solve the spectra of the a matrix $A$ in a given ensemble of random
matrix we introduce the generating function $\Gamma(\lambda)$
\be
{\Gamma}(\lambda)=\frac{1}{Z_{\phi}}\int \prod_{i=1}^{N}
\prod_{a=1}^n d\phi_i^a \prod_{i,a} 
\exp(\frac{i}{2}
\lambda \phi_i^a\phi_i^a) 
\prod_{<i,j>}\Avg{e^{{{-i}}\sum_a \phi_i^a A_{ij} \phi_j^a}}
\ee
with 
\be
Z_{\phi}=\int\prod_{i,a}d\phi_i^a e^{\sum_{i,a}\phi_i^a\phi_i^a}.
\ee
The spectral density is given by 
\be
{\rho(\lambda)}=\lim_{n\rightarrow 0}\frac{-2}{\pi nN}\mbox{Im}\frac{\partial}{\partial\lambda}\Avg{\Gamma(\lambda)}
\ee
\section{Spectra of adjacency matrix of sparse networks}
To solve the spectra of the adjacency matrix of a random complex
networks we introduce the generating function $\Gamma(\lambda)$
\be
{\Gamma}(\lambda)=\frac{1}{Z_{\phi}}\int \prod_{i=1}^{N}
\prod_{a=1}^n d\phi_i^a \prod_{i,a} 
\exp(\frac{i}{2}
\lambda \phi_i^a\phi_i^a) 
\prod_{<i,j>}\Avg{e^{{{-i}}\sum_a \phi_i^a a_{ij} \phi_j^a}}.
\ee
We assume that the support of our  matrix is a random 
uncorrelated network with given expected degree assigned to each node
of the network i.e. a realization of the
random hidden-variable model \cite{hv1,hv2,hv3,hv4,hv5}.
In particular we fix the expected degree distribution of each node $i$ 
of the undirected network  to be  $q_{i}$  and we assume
that the matrix elements $a_{i,j}$ are distributed following 
\be
P(a_{i,j})=\frac{q_{i}q_{j}}{\avg{q}N}\delta(a_{i,j}-1)+\left(1-
\frac{q_{i}q_{j}}{\avg{q}N}\right)\delta(a_{i,j}),
\label{HV}
\ee
for $i<j$ ($a_{ij}=a_{ji}$) and where  $\delta()$ indicates the Kronecker delta.
The partition function can then be average over the network ensembles
\bea
\Avg{\Gamma(\lambda)}&=&\frac{1}{Z_{\phi}}\int \prod_{i=1}^{N}
\prod_{a=1}^n d\phi_i^a \prod_{i,a} 
\exp(\frac{i}{2}
\lambda \phi_i^a\phi_i^a)\\ 
&&\times \exp\left\{-\frac{1}{2}\sum_{i,j}\frac{q_iq_j}{{\avg{q}N}}[1-\exp(i\sum_a \phi_i^a  
\phi_j^a)]+{\cal O}(N^{0})\right\}.\nonumber
\eea
We introduce the order parameters of the replicated variables on
sparse networks \cite{Monasson_order_pa}
\be
c_q(\vec{\phi})=\frac{1}{N_{q}} \sum_{i}\delta(q_i-q) \prod_a\delta(\phi_i^a-
\phi^a) 
\label{OP}
\ee
getting  for the partition function an expression of the type
\bea
\Avg{\Gamma(\lambda)}=\int {\cal D}c_q(\vec{\phi}) 
\exp[nN\Sigma(\{c_q(\vec{\phi})\})]\nonumber 
\eea
with
\begin{widetext}
\bea
n\Sigma &=&-\sum_q\int  d\vec{\phi}  p_q
{c}_q(\vec{\phi})\ln(c_q(\vec{\phi}))+i\sum_q p_q c_q(\vec{\phi})
\frac{1}{2}\lambda \sum_a\phi^a\phi^a\nonumber \\
&&- \int d\vec{\phi}\int  d\vec{\psi}\sum_{q q'}p_q p_{q'}\frac{1}{2}\frac{qq'}{\avg{q}} c_q(\vec{\phi})
c_{q'}({\vec{\psi}})(1-\exp(i\vec{\phi} \cdotp {\vec{ \psi}}))+{\cal O}(N^{-1}) \\
\eea
\end{widetext}

The saddle point equations for evaluating $\Sigma$ are given by
\bea
c_q(\vec{\phi})&=& \exp\left\{i\frac{\lambda}{2} \sum_a\phi^a
  \phi^a-q[1-\hat{c}(\vec{\phi})]\right\}\nonumber\\ 
\hat{c}(\vec{\phi})&=&\sum_{q'} \frac{q'p_{q'}}{\avg{q}} \int
  d\vec{\psi}\
  c_{q'}({\vec{\psi}})\exp(i\vec{\phi} \cdotp {\vec{ \psi}}) .
\label{sp.eq}
\eea

We assume that the solution of the saddle point equation is replica
symmetric, i.e. the distribution of the variables $\phi_a$ 
 conditioned
to a vector field ${\vec{x}}$ are identically equal distributed,
\be
c(\vec{\phi})=\int d\vec{x} P(\vec{x})\prod_{a=1}^n \Psi(\phi_a|\vec{x})
\ee

where $\Psi(\phi|\vec{x})$ are distribution functions of $\phi$ and
$P({\vec{x}})$ is a probability distribution  of the vector field
$\vec{x}$.
For the  function $\Psi(\phi|\vec{x})$ the exponential form is usually assumed 
in Ising models.
In our continuous variable case  for our quadratic problem, we assume
instead, as in \cite{Metabolico}, that  $\Psi(\phi|\vec{x})$  has a Gaussian form.
This assumption could be in general considered as an approximate solution
of the equations $(\ref{sp.eq})$.
Explicitly we   assume that the functions $c_q(\vec{\psi})$ ,  $\hat{c}(\vec{\psi})$
\bea
c_q(\vec{\phi})&=&\int dh_q P_q(h_q)\prod_a \exp\left [-\frac{1}{2}h_q
  \phi^a
\phi^a \right]
\left(\sqrt{\frac{h_q}{2\pi}}\right)^n\nonumber \\
\hat{c}(\vec{\phi})&=&\int  d\hat{h} \hat{P}(\hat{h})\prod_a \exp\left [-
\frac{1}{2}\hat{h}
  \phi^a \phi^a \right]. 
\label{gaus1.eq}
\eea

The saddle point equations $(\ref{sp.eq})$, taking into account the
expression for the order parameters $(\ref{gaus1.eq})$
 closes as in the problem studied in \cite{Metabolico} and can be written  as recursive equation for  $P_q(h_q), \hat{P}(\hat{h})$, i.e.
\begin{widetext}
\bea
P_q(h_q)&=&\sum_k
e^{-q} q^k\frac{1}{k!}\int_{\dots}\int \prod_{l=1}^k d\hat{h}^l \hat{P} 
(\hat{h}^l,\hat{m}^l)\delta\left(h_q-\sum_{l=1}^k \hat{h}^l-i\lambda\right)  
\nonumber \\
\hat{P}(\hat{h})&=& \sum_q \frac{q p_q}{\avg{q}} \int  dh_q  
\prod_i P_q(h_q)\delta\left(\hat{h}- \frac{1}{h_q}\right).   \label{recursive}
\eea
\end{widetext}
 
Once the distributions  $P_q(h_q)$ are found by the
population dynamics algorithm, then the  spectral density of the
network can be expressed as 
\be 
{\rho(\lambda)}=-\frac{1}{\pi}\sum_q p_q\int dh_q
P(h_q)\mbox{Im} \frac{i}{h_q}
\ee
Equations $\ref{recursive}$ can be solved as suggested in \cite{Metabolico}
 by a population dynamics algorithm.
 The action of the algorithm for finding $\hat{P}(\hat{h})$ is summarized in the following pseudocode
\\
{\bf algorithm } PopDyn($\{\hat{h}\}$)
{\bf begin}
{\bf do }
\begin{itemize}
\item select a random index $\alpha\in(1,M)$
\item choose a random $q$  with probability $q p_q$
\item draw {\it k} from a Poisson distribution ($e^{-q_i}q_i^k/k!$)
\item select $k$ indexes $\beta_1,\dots \beta_k\in\{1,\dots M\}$
\bea
\hat{h}^{\alpha}:&=&\frac{1}{i\lambda+\sum_{l=1}^kh_{\lambda}^{\beta_l}};\nonumber \\
\eea
\end{itemize}
{\bf while} (not converged)
{\bf return}
{\bf end}
\\
The effective medium  approximation $(EMA)$as found in
\cite{Doro_spectra} will be given by the solution of the population dynamics with $M=1$, i.e.
\be
\hat{h}_{EMA}=\sum_q \frac{qp_q}{\avg{q}}\frac{1}{i\lambda+q\hat{h}_{EMA}}.
\ee
The density in this approximation  take the form
\be
\rho(\lambda)=\sum_q p_q \mbox{Im} \frac{1}{\lambda-iq\hat{h}_{EMA}}
\ee

\section{Spectra of the Laplacian  }
The Laplacian  of a complex networks plays a crucial role in diffusion
process on the network and on the stability of many dynamical fixed
points \cite{May,Doro_lap,Syncrh}. The Laplacian  is defined in
terms of the adjacency matrix $a_{ij}$ of the network as the matrix 
of entries $L_{ij}=-a_{i,j}+\sum_{k} a_{ik} \delta_{ij}$.
For the Laplacian  matrix  the generating function $\Gamma(\lambda)$
takes the form,
\be
{\Gamma}(\lambda)=\frac{1}{Z_{\phi}}\int \prod_{i=1}^{N}
\prod_{a=1}^n d\phi_i^a \prod_{i,a} 
\exp(\frac{i}{2}
\lambda \phi_i^a\phi_i^a) 
\prod_{<i,j>}\Avg{e^{{{-i}}\sum_a \phi_i^a L_{ij} \phi_j^a}}.
\ee
Performing the average over the networks in  hidden variable ensemble
with fixed expected degree, Eq. $(\ref{HV})$,  we obtain  
\bea
\Avg{\Gamma(\lambda)}&=&\frac{1}{Z_{\phi}}\int \prod_{i=1}^{N} d\phi_i^a \prod_{i,a} \exp({i}
\frac{1}{2}\lambda \phi_i^a\phi_i^a)\nonumber\\ 
&=&\exp\left\{-\frac{1}{2}\sum_{i,j}\frac{q_iq_j}{{\avg{q}N}}\left[1-\exp\left(\frac{i}{2}(
    \phi_i^a  -\phi_j^a)^2\right)\right] +{\cal O}(N^0) \right\}.
\eea
Introducing the   order parameters
$c_q(\vec{\phi})$ defined in Eq. $(\ref{OP})$ we 
get  for the partition function
\bea
\Avg{\Gamma(\lambda)}=\int {\cal D}c_q(\vec{\phi}) 
\exp[nN\Sigma(\{c_q(\vec{\phi})\})]\nonumber 
\eea
with
\begin{widetext}
\bea
n\Sigma &=&-\sum_q\int  d\vec{\phi}  p_q
{c}_q(\vec{\phi})\ln(c_q(\vec{\phi}))+\frac{i}{2}\lambda\sum_q p_q c_q(\vec{\phi})
\sum_a\phi^a\phi^a\nonumber \\ &&- \frac{1}{2}\int d\vec{\phi}\int  d\vec{\psi}\sum_{q q'}p_q
p_{q'}\frac{1}{2}\frac{qq'}{\avg{q}} c_q(\vec{\phi})
c_{q'}({\vec{\psi}})\left\{1-\exp\left[\frac{i}{2}(\vec{\phi} - {\vec{ \psi}})^2\right]\right\}+{\cal 
O}(N^{-1}). \\
\label{spl}
\eea           
\end{widetext}
The saddle point equation determining the order parameter are 
\begin{widetext}
\bea
c_q(\vec{\phi})&=& \exp\left\{i\frac{\lambda}{2} \sum_a\phi^a
  \phi^a-q[1-\hat{c}(\vec{\phi})]\right\}\nonumber\\ 
\hat{c}(\vec{\phi})&=&\sum_{q'} \frac{q'p_{q'}}{\avg{q}} \int
  d\vec{\psi}\
  c_{q'}({\vec{\psi}})\exp\left[\frac{i}{2}(\vec{\phi} -\vec{ \psi})^2\right] .
\label{sp2.eq}
\eea
\end{widetext}
Again these equations can be solved with the Gaussian ansatz introduced
in \cite{Metabolico},  Eq. $(\ref{gaus1.eq})$,
\begin{widetext} 
\bea
P_q(h_q)&=&\sum_k
e^{-q} q^k\frac{1}{k!}\int_{\dots}\int \prod_{l=1}^k d\hat{h}^l \hat{P} 
(\hat{h}^l,\hat{m}^l)\delta\left(h_q-\sum_{l=1}^k \hat{h}^l-i\lambda\right)  
\nonumber \\
\hat{P}(\hat{h})&=& \sum_q \frac{q P_q}{\avg{q}}  \int dh_q  
\prod_i P_q(h_q)\delta\left(\hat{h}- \frac{1}{h_q-i}+i\right)   \label{recursivel}
\eea
\end{widetext}
Finally the  spectral density is given by
\be 
{\rho(\lambda)}=-\frac{1}{\pi}\sum_q p_q\int dh_q P(h_q)
\mbox{Im}\frac{i}{h_q}.
\ee
Equations $(\ref{spl})$ can again be solved by a population dynamics algorithm
 The action of the algorithm for finding $\hat{P}(\hat{h})$ is summarized in the following pseudocode
\\
{\bf algorithm } PopDyn($\{\hat{h}\}$)
{\bf begin}
{\bf do }
\begin{itemize}
\item select a random index $\alpha\in(1,M)$
\item choose a random $q$  with probability $q p_q$
\item draw {\it k} from a Poisson distribution ($e^{-q_i}q_i^k/k!$)
\item select $k$ indexes $\beta_1,\dots \beta_k\in\{1,\dots M\}$
\bea
\hat{h}^{\alpha}:&=&\frac{1}{i(\lambda-1)+\sum_{l=1}^kh_{\lambda}^{\beta_l}}-i;\nonumber \\
\eea
\end{itemize}
{\bf while} (not converged)
{\bf return}
{\bf end}
Once the distribution of $\hat{h}$ is found from the first equation
of $(\ref{recursivel})$ it is strait-forward to calculate the
distributions for $P_q(h_q)$.
The effective medium  approximation $(EMA)$ as found in
 will be given by the solution of the population dynamics with $M=1$, i.e.
\be
\hat{h}_{EMA}=\sum_q \frac{qp_q}{\avg{q}}\frac{1}{i(\lambda-1)+q\hat{h}_{EMA}}-i.
\ee
The density in this approximation  take the form
\be
\rho(\lambda)=\sum_q p_q \mbox{Im} \frac{1}{\lambda-iq\hat{h}_{EMA}}
\ee
\section{Weighted networks}
The over-mentioned  results can be extended to weighted networks with weight degree
correlations. The correlation between the weight of the links $A_{ij}$
ending to a node $i$ and the degree of the node $i$ have been observed in
different networks \cite{PNAS} and can also be explained by growing
network models \cite{weig}.  
A network ensemble with weight degree correlations  can be formulated
by assuming that the weight of a link between node $i$ and node $j$,
if present, has a value $w_{ij}=C(q_i q_j)^{\theta}$  where $q_i$ and
$q_j$ are the expected conductivities of node $i$ and $j$ and $C,
\theta$ are two parameters specifying the ensemble under consideration.
Therefore in the following we will consider the symmetrix matrix $w_{ij}$  with
distribution of the matrix elements given by  
\be
P(w_{i,j})=\frac{q_{i}q_{j}}{\avg{q}N}\delta(w_{i,j}-C(q_i q_j)^{\theta})+\left(1-
\frac{q_{i}q_{j}}{\avg{q}N}\right)\delta(w_{i,j}).
\label{Pw}
\ee
for $i<j$ and $w_{ij}=w_{ji}$
The generating function  $\Gamma(\lambda)$ for this ensemble of
networks is given by 
\be
{\Gamma}(\lambda)=\frac{1}{Z_{\phi}}\int \prod_{i=1}^{N}
\prod_{a=1}^n d\phi_i^a \prod_{i,a} 
\exp(\frac{i}{2}
\lambda \phi_i^a\phi_i^a) 
\prod_{<i,j>}\Avg{e^{{{-i}}\sum_a \phi_i^a w_{ij} \phi_j^a}}.
\ee
with its average over the distribution $(\ref{Pw})$ taking the usual  form 
\bea
\Avg{\Gamma(\lambda)}=\int {\cal D}c_q(\vec{\phi}) 
\exp[nN\Sigma(\{c_q(\vec{\phi})\})]. \nonumber 
\eea
with  $c_q(\vec{\phi})$ given by $(\ref{OP})$ and 
\begin{widetext}
\bea
n\Sigma &=&-\sum_q\int  d\vec{\phi}  p_q
{c}_q(\vec{\phi})\ln(c_q(\vec{\phi}))+\frac{i}{2}\lambda\sum_q p_q c_q(\vec{\phi})
\sum_a\phi^a\phi^a\nonumber \\ &&- \frac{1}{2}\int d\vec{\phi}\int  d\vec{\psi}\sum_{q q'}p_q
p_{q'}\frac{1}{2}\frac{qq'}{\avg{q}} c_q(\vec{\phi})
c_{q'}({\vec{\psi}})[1-\exp({i}(C(qq')^{\theta}\vec{\phi}\cdot\vec{ \psi})^2)]+{\cal O}(N^{-1}) 
\eea
\end{widetext}
The saddle point equation to be solved are 

\bea
c_q(\vec{\phi})&=& \exp\left\{\frac{1}{2}\lambda \sum_a\phi^a
  \phi^a-q[1-\hat{c}_q(\vec{\phi})]\right\}\\ 
\hat{c}_q(\vec{\phi})&=&\sum_{q'} p_{q'}\frac{q'}{\avg{q}}
\int  c_{q'}({\vec{\psi}})
  d\vec{\psi} \exp\left[iC(qq')^{\theta}\vec{\phi} \cdot \vec{ \psi}\right] .\nonumber
\label{sp3.eq}
\eea
The recursive equations to be solved ad the saddle point are
\begin{widetext}
\bea
P_q(h_q)&=&\sum_k
e^{-q} q^k\frac{1}{k!}\int_{\dots}\int \prod_{l=1}^k d\hat{h}^l_q \hat{P}_q 
(\hat{h}^l_q)\delta\left(h_q-\sum_{l=1}^k \hat{h}^l_q-i\lambda\right)  
\nonumber \\
\hat{P}_q(\hat{h}_q)&=& \sum_{q'} \frac{q' P_{q'}}{\avg{q}}  \int dh_{q'}  
 P_{q'}(h_{q'})\delta\left(\hat{h}_q- \frac{C^2(qq')^{2\theta}}{h_{q'}}\right)   \label{recursivew}
\eea
\end{widetext}
The spectral density is given by
\be 
{\rho(\lambda)}=-\frac{1}{\pi}\sum_q p_q\int dh_q P(h_q)\mbox{Im} \frac{i}{h_q}
\ee

 The equations $(\ref{recursivew})$  can be solved by a population-dynamical algorithm.

 The action of the algorithm is summarized in the following pseudo
 code
\\
\\
{\bf algorithm } PopDyn($\{ \hat{h}_q\}$)
{\bf begin}
{\bf do }
\begin{itemize}
\item select a random $q$ and a random  index $\alpha\in(1,M)$
\item choose a random $q'$  with probability $q' p_{q'}$
\item draw {\it k} from a Poisson distribution ($e^{-q'}q'^k/k!$)
\item select $k$ indexes $\beta_1,\dots \beta_k\in\{1,\dots M\}$
\bea
\hat{h}^{\alpha}_q:&=&\frac{C^2(qq')^{\theta}}{i\lambda+ \sum_{l=1}^k\hat{h}_{q'}^{\beta_l}};\nonumber \\
\eea
\end{itemize}
{\bf while} (not converged)
{\bf return}
{\bf end}
The equivalent of the effective medium approximation are the following
equation for $\hat{h}_q^{(EMA)}$
\be
\hat{h}_q^{(EMA)}=\sum_{q'}\frac{q'p_{q'}}{\avg{q}}\frac{C^2(qq')^{\theta}}{i\lambda+q'\hat{h}_{q'}^{(EMA)}}
\ee
and the spectral density is given by  
\be
\rho(\lambda)=\sum_q p_q\mbox{Im}\frac{1}{\lambda-iq\hat{h}_q^{(EMA)}}
\ee
In conclusion we have provided a solution for the non linear integral equations 
for the spectra density in random matrix theory of the spectra of
sparse random matrices introducing the order parameter as product of
Gaussian distributions, the applications of this approach will be
relevant in many fields and stability of stationary state in dynamical
system defined on complex networks.

After this work was completed we become aware of similar findings
obtained by R. Kuehn \cite{reimer}.

\end{document}